# Interacting CMEs and their associated flare and SEP activities

A.Shanmugaraju[1] and S.Prasanna Subramanian[2]

[1]Arul Anandar College, Karumathur- 625 514, Madurai Dist., India.

and

[2]M.K.University College, Aundipatti, Theni Dist., India.

## Abstract

We have analyzed a set of 25 interacting events which are associated with the DH type II bursts. These events are selected from the Coronal Mass Ejections (CMEs) observed during the period 1997 – 2010 in SOHO/LASCO and DH type IIs observed in Wind/WAVES. Their pre and primary CMEs from nearby active regions are identified using SOHO/LASCO and EIT images and their height – time diagrams. Their interacting time and height are obtained, and their associated activities, such as, flares and Solar Energetic Particles( >10 pfu) are also investigated. Results from the analysis are: primary CMEs are much faster than the pre-CMEs, their X-ray flares are also stronger (X and M class) compared to the flares (C and M class) of pre-CMEs. Most of the events (22/25) occurred during the period 2000 – 2006. From the observed width and speed of pre and primary CMEs, it is found that the pre-CMEs are found to be less energetic than the primary CMEs. While the primary CMEs are tracked up to the end of LASCO field of view (30Rs), most of the pre-CMEs can be tracked up to <26 Rs. The SEP intensity is found to be related with the integrated flux of X-ray flares associated with the primary CMEs for nine events originating from the western region.

# 1 Introduction

Coronal Mass Ejection (CME) is the transient ejection of plasma and magnetic fields from the sun's corona into the interplanetary space detected by coronagraphs. CMEs are responsible for large geomagnetic storms and Solar Energetic Particle emission (SEP). Type II radio bursts is an important tool to understand the physical phenomena in the corona of the solar atmosphere and in the interplanetary (IP) medium. CMEs are often associated with Type II radio bursts that are also indicators of coronal and interplanetary shocks. The collision between a slow pre-CME and fast primary CME was reported by Gopalswamy et al. (2001a) and they found radio signatures due to interaction. On the other hand, Lugaz et al.(2005) studied a simulation of interaction between two identical CMEs having same speed in the interplanetary medium. Their simulation results reproduced some general features observed in satellite data for multiple magnetic clouds. Interacting CMEs associated with Type II radio burst cause solar energetic particles as listed by Gopalswamy et al.(2001b). Type II radio burst occurring at all wavelengths from metric to kilometric are associated with most energetic CMEs (Gopalswamy et al.,2005). Long wavelength radio emission in decameter-hectometric (DH) wavelength from 1kHz to 14 MHz frequency range is important to understand the effects of CME in the outer corona and in the IP medium. Also, the radio signatures due to CME interaction are observed in the DH range of spectrum.

Gopalswamy et al.(2002) reported a high correlation between interacting CMEs and SEP association using a set of events during 1996-2001. The emission of non-thermal electrons in the interplanetary medium happens due to interaction between the CMEs. Gopalswamy et al.(2004) emphasized more research on the interacting CMEs because it is important in the concept of space-weather. Recently, Oliverous et al.(2012) have studied an interacting event on 2010 August 01, and suggested that Type II burst radio emission produced is related to CME interaction. The same event is analyzed by Temmer et al.(2012). In their kinematic study, they suggest that the increase in magnetic tension and pressure when primary CME bends and compress the magnetic field lines of pre-CME, increases the efficiency of drag.

Recently, Prasanna Subramanian and Shanmugaraju (2013) analyzed a set of 15 interacting CMEs from the list of events observed during 1997-2002 by Manoharan et al.(2004). They found existence of relations between interaction height and time delay, and interaction height and frequency of radio emissions. We have extended their study to a wider period during 1997-2010 to identify more events and to complement their results. In this paper, a larger set of 25 interaction events is identified during this period. In addition to the properties of pre and primary CMEs, their associated flare activities are also analyzed.

## 2    Data Selection

A set of 345 Type II radio bursts in decameter-hectometric (DH) range (listed in Wind/WAVES Type II catalog (http://cdaw.gsfc.nasa.gov/CME_list/radio/waves_type2.html) between 1-14 MHz associated with major flares and CMEs observed during the period 1997-2010 is considered for the present study. The preliminary data of flares and CMEs associated with these events were obtained from the same catalog. Full data corresponding to CMEs are taken from the online catalog http://.cdaw.gsfc.gov/CME list (Yashiro et al., 2004) and X-ray flare data(start time, duration, location, X-ray integrated flux and flare class) are obtained from the website http://www.ngdc.noaa.gov/stp/spaceweather.html. The flare-CME association has been examined using the temporal and spatial relationship as given in Harrison (1986, 1995), Jing et al.(2003), Yashiro and Gopalswamy(2008) and Yan et al.(2011). The flare locations and peak flux are also confirmed by both EIT images of SOHO and SDO flare locator images (http://www.solarmonitor.org/). The CMEs within one hour duration after flare onset are carefully examined whether they are ejected from that flare location. The CMEs ejected from the same locations are obtained from LASCO coronagraph images and their X-ray flares are obtained from GOES X ray plots.

First, the CMEs causing the IP type II bursts were identified from LASCO C2 and C3 coronagraph observations. For an interacting event, a slow moving CME called pre-CME got interaction with the next or after ejected fast CME, called primary CME. Height-time plots of CMEs are drawn together to see if their trajectories of pre and primary CMEs are intersecting with each other. The height-time plot is extended up to 60 Rs to see any interaction after the LASCO field of view. The interacting CMEs were confirmed with LASCO images of SOHO. Finally, we have selected a set of 25 interacting events based on the following selection criteria: (i) the pre-CMEs ejected before 9 hours from the nearby active region in the same quadrant were identified. (ii) only close events having separation in position angle less than 90° between the events from nearby active region were considered. (iii) primary CMEs associated with DH Type II radio bursts and X-ray flares were considered for the study. (iv) pre-CMEs should be associated with flares and having minimum width 30degreeswere only considered.

The other parameters of CMEs and associated activities are obtained as follows.The time delay of an interacting event is calculated between the onset times of pre and primary CMEs. Some events are found to be associated with major SEP proton events (>10 pfu.) From GOES proton data, we identified 10 major SEP events. SEP data are taken from the website www.umbra.nascom.nasa.gov/sep. The Final Observed Distance(FOD) of all the pre and primary CME events are obtained. If there were more than one preceding CMEs, we considered the pre-CME having more width and less time interval, and it should interact with primary CME as seen in coronagraph observation. All the data (date, time of first detection in LASCO, speed, central position angle (CPA), width and FOD) of pre and primary CMEs are listed in column 2 and 3, respectively in Table 1.

**Table 1:** Data corresponding to all the pre and primary CMEs observed during the period 1997 – 2010

| No. | Pre CME | | | | | | Primary CME | | | | | |
|---|---|---|---|---|---|---|---|---|---|---|---|---|
| | Date | Time | Speed | CPA | Width | FOD | Date | Time | Speed | CPA | Width | FOD |
| | | hh:mm | km/s | deg | deg | Rs | | hh:mm | km/s | deg | deg | Rs |
| 1 | 1997 Nov 06 | 4:20 | 307 | 263 | 59 | 12 | 1997 Nov 06 | 12:10 | 1556 | Halo | 360 | 25 |
| 2 | 2000 July 22 | 8:30 | 204 | 308 | 20 | 5 | 2000 July 22 | 11:54 | 1230 | 259 | 229 | 19 |
| 3 | 2001 Jan 20 | 19:31 | 839 | Halo | 360 | 25 | 2001 Jan 20 | 21:30 | 1507 | Halo | 360 | 28 |
| 4 | 2001 Apr 02 | 10:06 | 203 | 278 | 36 | 3 | 2001 Apr 02 | 11:26 | 992 | 270 | 80 | 24 |
| 5 | 2002 Apr 14 | 4:06 | 279 | 311 | 42 | 4 | 2002 Apr 14 | 7:50 | 757 | 323 | 76 | 21 |
| 6 | 2003 Mar 18 | 7:31 | 619 | 204 | 71 | 4 | 2003 Mar 18 | 12:30 | 1601 | 263 | 209 | 26 |
| 7 | 2003 May 27 | 23:50 | 964 | Halo | 360 | 17 | 2003 May 28 | 0:50 | 1366 | Halo | 360 | 29 |
| 8 | 2003 Nov 18 | 8:06 | 1223 | 144 | 104 | 26 | 2003 Nov 18 | 8:50 | 1660 | Halo | 360 | 27 |
| 9 | 2003 Dec 2 | 8:26 | 240 | 231 | 41 | 4 | 2003 Dec 2 | 10:50 | 1393 | 261 | 150 | 24 |
| 10 | 2004 July 23 | 17:54 | 569 | 260 | 142 | 13 | 2004 July 23 | 19:31 | 874 | 209 | 100 | 23 |
| 11 | 2004 July 25 | 14:30 | 450 | 228 | 45 | 4 | 2004 July 25 | 14:54 | 1333 | Halo | 360 | 21 |
| 12 | 2004 Nov 06 | 01:31 | 818 | Halo | 360 | 8 | 2004 Nov 06 | 2:06 | 1111 | 351 | 214 | 27 |
| 13 | 2004 Nov 07 | 14:30 | 226 | 286 | 100 | 5 | 2004 Nov 07 | 16:54 | 1759 | Halo | 360 | 22 |
| 14 | 2004 Dec 29 | 9:21 | 345 | 105 | 36 | 14 | 2004 Dec 29 | 16:45 | 774 | 71 | 140 | 25 |
| 15 | 2004 Dec 30 | 17:54 | 231 | 87 | 54 | 6 | 2004 Dec 30 | 22:30 | 1035 | Halo | 360 | 22 |
| 16 | 2005 Jan 17 | 9:30 | 2094 | Halo | 360 | 24 | 2005 Jan 17 | 9:54 | 2547 | Halo | 360 | 21 |
| 17 | 2005 Jan 20 | 3:54 | 205 | 237 | 35 | 4 | 2005 Jan 20 | 6:54 | 882 | Halo | 360 | 13 |
| 18 | 2005 June 03 | 3:32 | 247 | 126 | 128 | 12 | 2005 June 03 | 12:32 | 1679 | Halo | 360 | 23 |
| 19 | 2005 July 07 | 13:26 | 432 | 118 | 165 | 16 | 2005 July 07 | 17:06 | 683 | Halo | 360 | 13 |
| 20 | 2005 July 13 | 12:54 | 471 | 275 | 49 | 8 | 2005 July 13 | 14:30 | 1423 | Halo | 360 | 26 |
| 21 | 2005 July 14 | 7:54 | 752 | 266 | 103 | 26 | 2005 July 14 | 10:54 | 2115 | Halo | 360 | 14 |
| 22 | 2005 July 30 | 5:57 | NA* | 75 | 30 | 2* | 2005 July 30 | 6:50 | 1968 | Halo | 360 | 26 |
| 23 | 2005 Sep 13 | 13:24 | 229 | 36 | 40 | 11 | 2005 Sep 13 | 20:00 | 1866 | Halo | 360 | 22 |
| 24 | 2007 June 03 | 6:54 | 208 | 84 | 36 | 4 | 2007 June 03 | 09:54 | 467 | 86 | 71 | 8 |
| 25 | 2010 Aug 18 | 0:24 | 403 | 298 | 88 | 19 | 2010 Aug 18 | 5:48 | 1471 | 255 | 184 | 25 |

**Table 2:** Data of interaction height, time, time delay and details of x-ray flare associated with the primary CMEs.

| No. | Date | CME Time | Speed (km/s) | Interaction height (Rs) | Interaction time | Time delay (min) | X-ray Class | Active Region of flare | flare location | Duration of flare (min) | x-ray integrated flux ($10^{-2}$) J/m$^{-2}$ | SEP Intensity (pfu) |
|---|---|---|---|---|---|---|---|---|---|---|---|---|
| 1 | 1997 Nov 06 | 12:10 | 1556 | 21 | 14:24 | 549 | X9.4 | **8100** | S18W63 | 12 | 36 | 490 |
| 2 | 2000 July 22 | 11:54 | 1230 | 8 | 12:30 | 272 | M3.7 | **9085** | N14W56 | 26 | 7 | 17 |
| 3 | 2001 Jan 20 | 21:30 | 1507 | 24 | 0:00 | 139 | M7.7 | **9313** | S07E46 | 49 | 7.2 | - |
| 4 | 2001 Apr 02 | 11:26 | 992 | 6 | 12:24 | 144 | X.1 | **9393** | N20W70 | 67 | 30 | - |
| 5 | 2002 Apr 14 | 7:50 | 757 | 10 | 9:42 | 234 | C9.6 | **9893** | N19W57 | 16 | 0.57 | - |
| 6 | 2003 Mar 18 | 12:30 | 1601 | 27 | 15:24 | 301 | X1.5 | **10314** | S15W46 | 39 | 5.1 | - |
| 7 | 2003 May 28 | 0:50 | 1366 | 28 | 4:24 | 85 | X3.6 | **10365** | S07W17 | 33 | 28 | 121 |
| 8 | 2003 Nov 18 | 8:50 | 1660 | 8 | 9:15 | 28 | M3.9 | **10501** | N00E18 | 47 | 8.4 | - |
| 9 | 2003 Dec 02 | 10:50 | 1393 | 8 | 11:52 | 185 | C7.2 | **10508** | S14W7 0 | 14 | 0.51 | 86 |
| 10 | 2004 July 23 | 19:31 | 874 | 18 | 22:42 | 121 | C4.1 | **10652** | N04W05 | 20 | 0.17 | - |
| 11 | 2004 July 25 | 14:54 | 1333 | 7 | 15:42 | 67 | M1.1 | **10652** | N08W33 | 144 | 6.5 | 2086 |
| 12 | 2004 Nov 06 | 2:06 | 1111 | 19 | 4:36 | 52 | M3.6 | **10696** | N09E05 | 28 | 5.5 | - |
| 13 | 2004 Nov 07 | 16:54 | 1759 | 5 | 16:36 | 330 | X2.0 | **10696** | N09W17 | 33 | 20 | 495 |
| 14 | 2004 Dec 29 | 16:45 | 774 | 26 | 21:00 | 456 | M2.3 | **10715** | N04E62 | 41 | 1.8 | - |
| 15 | 2004 Dec 30 | 22:30 | 1035 | 9 | 23:24 | 329 | M4.2 | **10715** | N04E46 | 26 | 3.6 | - |
| 16 | 2005 Jan 17 | 9:54 | 2547 | * | * | 36 | X3.8 | **10720** | N15W25 | 188 | 84 | 5040 |
| 17 | 2005 Jan 20 | 6:54 | 882 | 9 | 7:48 | 230 | X7.1 | **10720** | N14W61 | 50 | 130 | 1680 |
| 18 | 2005 June 03 | 12:32 | 1679 | 15 | 13:30 | 604 | M1.0 | **10772** | N15E90 | 54 | 1.8 | - |
| 19 | 2005 July 07 | 17:06 | 683 | 23 | 22:30 | 232 | M4.9 | **10786** | N09E03 | 33 | 5.3 | - |
| 20 | 2005 July 13 | 14:30 | 1423 | 8 | 15:18 | 127 | M5.0 | **10786** | N11W90 | 98 | 20 | 134 |
| 21 | 2005 July 14 | 10:54 | 2115 | 23 | 12:48 | 205 | X1.2 | **10786** | N11W90 | 73 | 39 | - |
| 22 | 2005 July 30 | 06:50 | 1968 | * | * | 26 | X1.3 | **10792** | N12E60 | 44 | 23 | - |
| 23 | 2005 Sep 13 | 20:00 | 1866 | 12 | 21:00 | 509 | X1.5 | **10808** | S16E39 | 98 | 55 | - |
| 24 | 2007 June 03 | 09:54 | 469 | 14 | 15:00 | 211 | C5.3 | **10960** | S08E67 | 8 | 0.12 | - |
| 25 | 2010 Aug 08 | 05:48 | 1471 | 16 | 7:12 | 403 | C4.5 | **11099** | N18W88 | 126 | 2.3 | - |

*- not possible to measure

**Table 3**:   Details of  X-ray flare class  of Pre-CMEs.

| No. | Date | CME Time | X-ray Class | Active Region of flare | flare location | Duration of flare (min) | x-ray integrated flux ($10^{-2}$ )J/m$^{-2}$ |
|---|---|---|---|---|---|---|---|
| 1 | 1997 Nov 06 | 4:20 | C1.9 | **8100** | **S15W56** | 29 | 0.28 |
| 2 | 2000 July 22 | 8:30 | M1.0 | **9090** | N12W05 | 32 | 1.4 |
| 3 | 2001 Jan 20 | 19:31 | M1.2 | **9313** | S07E40 | 26 | 1.2 |
| 4 | 2001 Apr 02 | 10:06 | M1.9 | **9393** | N17W60 | 14 | 1.3 |
| 5 | 2002 Apr 14 | 4:06 | M1.4 | **9893** | N22W45 | 48 | 3.2 |
| 6 | 2003 Mar 18 | 7:31 | C2.1 | **10314** | S15W45 | 32 | 0.24 |
| 7 | 2003 May 27 | 23:50 | X1.3 | **10365** | S07W17 | 17 | 7.1 |
| 8 | 2003 Nov 18 | 8:06 | M2.3 | **10501** | N00E18 | 39 | 5.1 |
| 9 | 2003 Dec 02 | 8:26 | C3.9 | **10508** | SW | 20 | 0.37 |
| 10 | 2004 July 23 | 17:54 | M2.2 | **10652** | N04W08 | 32 | 1.5 |
| 11 | 2004 July 25 | 14:30 | M2.2 | **10652** | N04W30 | 18 | 1.3 |
| 12 | 2004 Nov 06 | 1:31 | M5.9 | **10696** | NE | 26 | 8.5 |
| 13 | 2004 Nov 07 | 14:30 | C7.0 | **10696** | N08W14 | 22 | 0.63 |
| 14 | 2004 Dec 29 | 9:21 | C2.6 | **10713** | NE | 119 | 1.2 |
| 15 | 2004 Dec 30 | 17:54 | B5.1 | **10715** | N03E49 | 7 | 0.017 |
| 16 | 2005 Jan 17 | 9:30 | X2.0 | **NA** | **N13W19** | NA | NA |
| 17 | 2005 Jan 20 | 3:54 | C4.8 | **10720** | N19W58 | 15 | 0.27 |
| 18 | 2005 June 03 | 3:32 | C3.1 | **10772** | S17E21 | 45 | 0.45 |
| 19 | 2005 July 07 | 13:26 | C2.6 | **10789** | N17E48 | 37 | 0.48 |
| 20 | 2005 July 13 | 12:54 | M3.2 | **10786** | N08W79 | 21 | 1.3 |
| 21 | 2005 July 14 | 7:54 | M9.1 | **10786** | N09W90 | 106 | 8.4 |
| 22 | 2005 July 30 | 5:57 | C9.4 | **10792** | N11E53 | 33 | 1.1 |
| 23 | 2005 Sep 13 | 13:54 | C4.5 | **10808** | SE | 42 | 0.95 |
| 24 | 2007 June 03 | 6:54 | M4.5 | **10960** | S06E63 | 7 | 0.93 |
| 25 | 2010 Aug 18 | 00:54 | C1.5 | **11099** | NW | 77 | 0.54 |

## 3     Results

For example, a C class flare  observed from the active region 10696   in the location N08W14 is shown in the left panel of Figure 1a.   Right panel shows a pre-CME ejected at 14:30UT on 07 November 2004 with a speed 226 km/s and width 100$^o$ at position angle 286°.  An X class flare is associated with the primary CME shown in left panel of Figure 1b from the active region 10696 in the location N09W17.  A halo  primary CME  at 16:54 UT with speed 1759 km/s and width 360° is shown in the Fig. 1b (right panel). Around 17:00 UT, the primary CME interacted with the pre-CME. Radio type II bursts observed in the Wind/WAVES dynamic spectrum below 14 MHz corresponding to the primary CME is shown in Fig.2. It is reported that the frequency range of this DH type II is 14 MHz – 60 kHz during the interval 16:25 UT on 7 November to 20:00 UT of next day. Fig.3 shows X-ray flare profile associated with the pre and primary CMEs. It seems that the flare associated with pre-CME have peak X-ray flux C7.0 at 14:06 UT and the flare associated with primary CME have peak X-ray flux X2.0 at 16:06 UT . Fig.4 shows the height-time diagram of both the CMEs in which the interaction is seen ~ 17:00 UT around a height of 5 to 7 Rs. At this time of interaction, the position of leading edges of these two CMEs are reported as pre-CME = 6 Rs and primary CME = 7Rs.

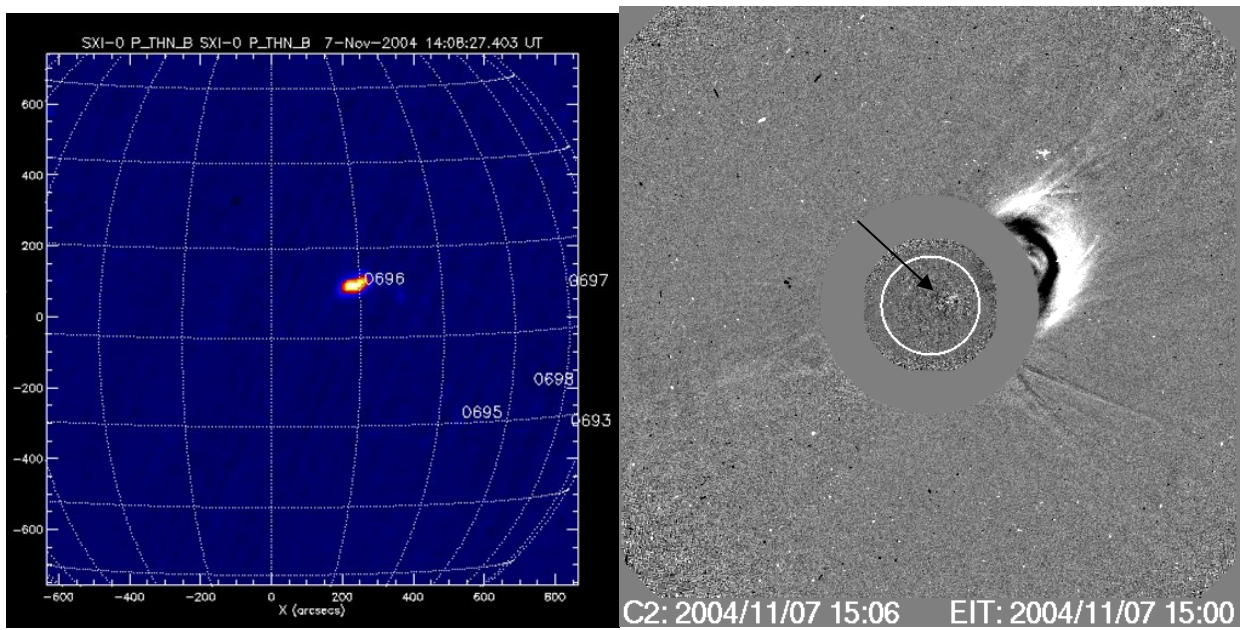

Fig.1a       C class flare ejected from the active region 10696 at the location N08W14 is shown (SXI image) in the left panel. The pre-CME ejected at 14:30UT (speed=226 km/s, CPA= 286°, width=100°) is shown (C2 coronagraph picture) in the right panel. The arrow mark indicates the active region of flare at which the pre-CME ejected.

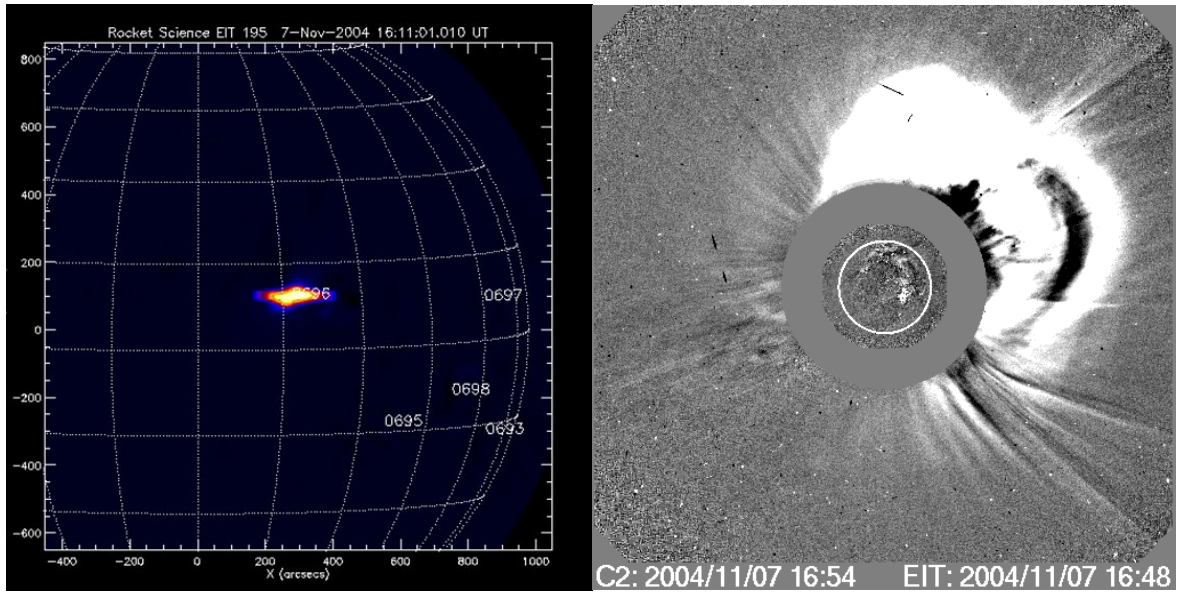

Fig. 1b  X class flare ejected from the active region 10696 at the location N09W17 is shown in the left panel. The primary CME ejected at 16:54UT (speed=1759 km/s, width=360°) is shown (C2 coronagraph picture) in the right panel. The mean position angle of the pre- CME is 298° and primary CME is 360°. Interaction of the two CMEs occurred ~17:00 UT around a height of ~5 Rs is seen on the right panel.

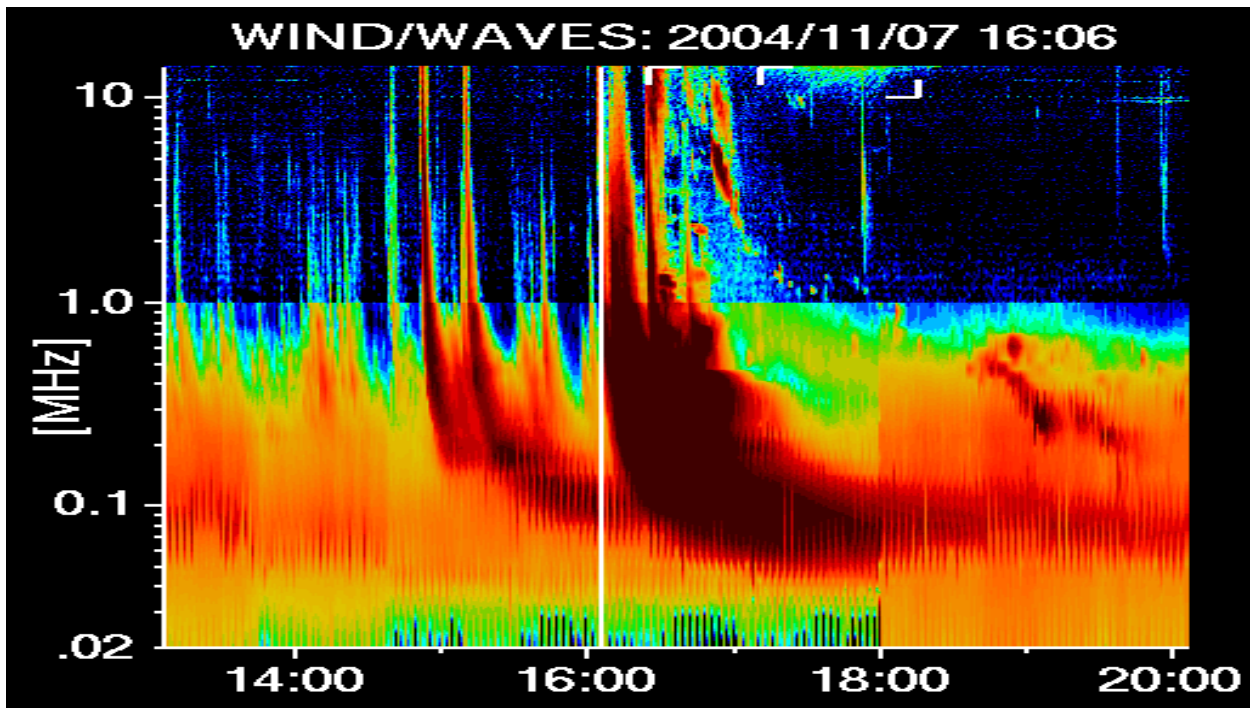

Fig. 2  DH type II burst recorded by Wind/Waves on 7 November 2004. The radio burst started around 14 MHz at 16:25 UT.

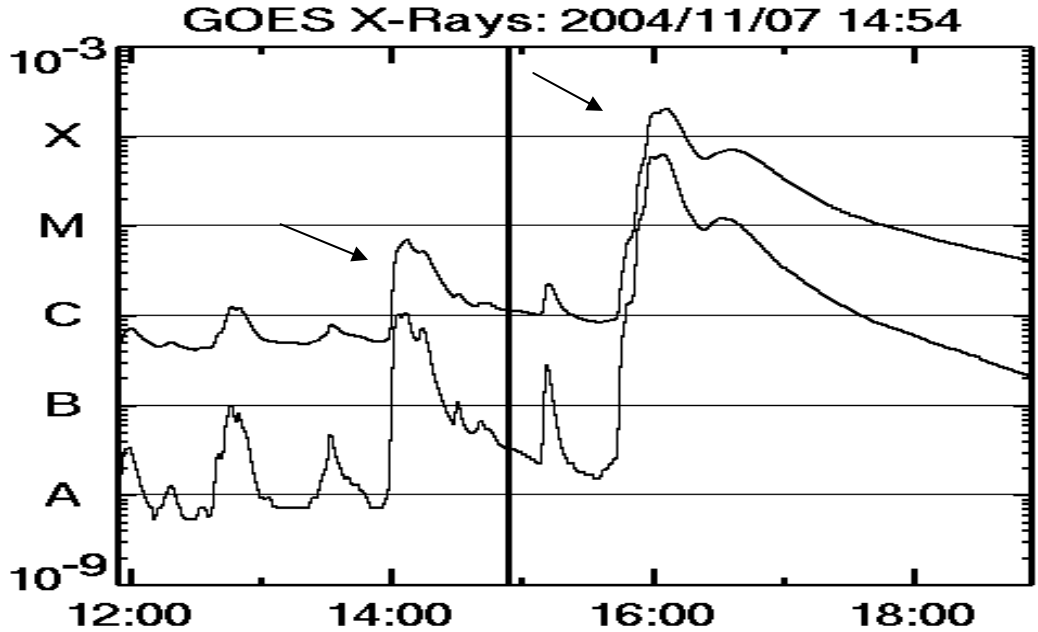

Fig. 3  GOES X-Ray flare profiles on 07 November 2004. The left arrow shows a C-class flare assocated with pre-CME and the right arrow shows an X-class flare associated with the primary CME. The vertical line denotes the time of observation 14:54UT.

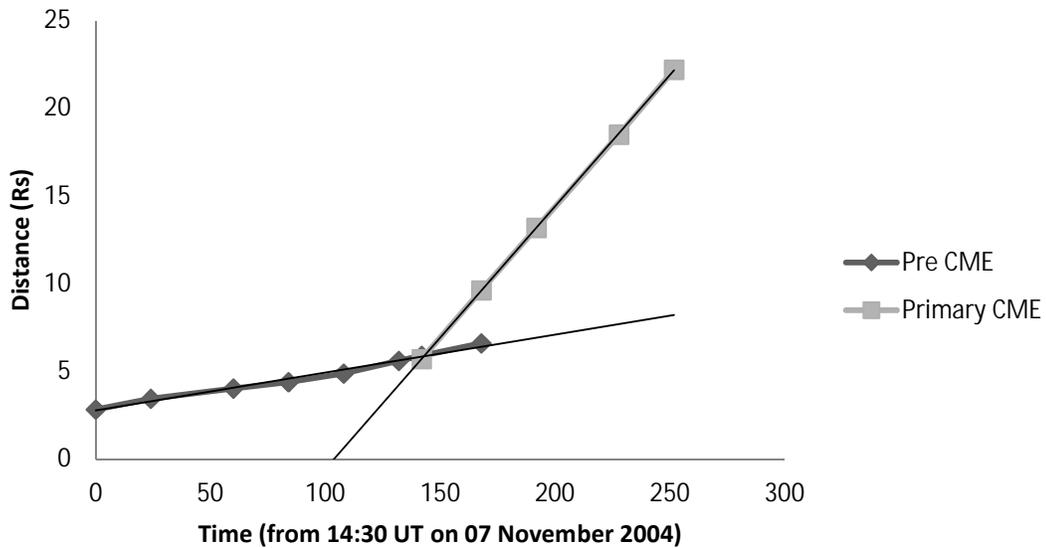

Fig. 4  Height-time diagram of pre and primary CMEs drawn together to show the interaction on 07 November 2004. The interaction of two CMEs occurred ~17:00 UT around a height of ~5 Rs.

The interaction height and interaction time are estimated from the height-time plots and they are listed in Table 2. Column 2-4 represent the date, time of first detection and linear speed of primary CMEs. The interaction height, time and delay time between the onsets of pre and primary CMEs are given in column **5-7**. Details of X-ray flares (class, Active Region number, location, duration and integrated flux) associated with the primary CMEs are given in column 8-12). The SEP intensity data (> 10pfu) corresponding primary CMEs is given the last column. A distribution plot of interaction heights is shown in Figure 5. The range of interaction height is between 5 – 28Rs and mean interaction height of the events is found to be 15Rs. It is nearly similar to the mean interaction height of CMEs associated with SEP events (21 Rs) observed by Gopalswamy (2002) and mean interaction height of CMEs with DH type II events (18 Rs) determined by Prasanna and Shanmugaraju(2013). Also, delay between the onsets of pre and primary CMEs is calculated. The delay lies in the range 26-604 minutes with a mean value of 235mins.

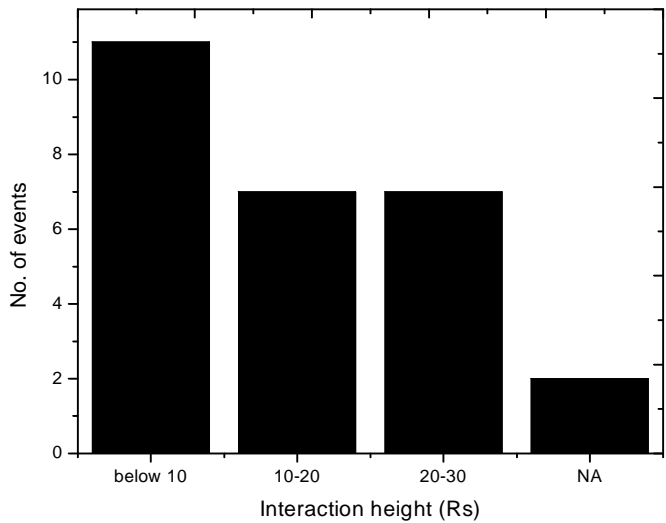

Fig. 5 Histogram plot shows the distribution of interaction height determined from the over-plotted height-time plots of pre and primary CMEs.

Similarly, the properties of all the pre and primary CMEs and their associated activities are obtained and their properties are analyzed further statistically. For example,

Figure 6 shows the distribution of speeds of pre and primary CMEs. It is seen that most of pre-CMEs have speed less than 1000km/s, whereas, many of the primary CMEs have speeds in the range 500 – 2000 km/s. The mean speed of the primary CMEs is 1361 km/s which is found to be more than twice the mean speed 523 km/s of pre-CMEs. This is similar to the mean speed of pre and primary CMEs reported by Gopalswamy et al.(2002) and Prasanna and Shanmugaraju (2013). Similarly the final observed distance (FOD) of primary CME is nearly twice the FOD of pre-CMEs. Most of the primary CMEs are halo CMEs, whereas, pre-CMEs have a lower mean width of 121 degrees.

Quite recently, Ding et al.(2013) reported the twin-CME scenario and large solar energetic particle events (SEPs) in solar cycle 23 using a set of fast CMEs (v > 900 km/s and width > 60 degrees) . They found four groups of events: (i) twin and (ii) single CMEs associated with large SEPs, (iii) twin and (iv) single CMEs not associated with large SEPs. Of 59 large SEP events, 43 and 16 events are found by them to be associated with twin and single CMEs, respectively. Also they reported that not all twin CMEs are associated with large SEP events. They suggested that the studies of the role of preceding flares are necessary to better understand the twin-CME scenario. Among 25events in the present study, 9 events are listed in twin-CME category of Ding et al.(2013).

The X-ray flares associated with the primary CMEs are obtained from the Wind/WAVEs catalog and those associated with the pre-CMEs are obtained from the flare catalog by following certain time window of +/- one hour from the first appearance time of pre-CMEs. They are listed in Tables 2 and 3 for primary CMEs and pre-CMEs, respectively, along with their properties like flare class, duration and X-ray integrated flux. The number of C, M and X-class flares associated with the primary CMEs are 5, 10 and 12, respectively. While the flares associated with the primary CMEs are predominantly M and X-class flares, the flares associated with pre-CMEs are mostly C and M-class flares. This is in agreement with the recent results of Ding et al.(2013) where they showed that the mean strength of X-ray flares associated with pre-CMEs (C9.1) and primary CMEs (M9.6). The distributions of number of flares in each class are given in Figure 7. Flare duration is estimated for both groups of flares associated with pre and primary CMEs. The mean value is found to be 55 min for primary CMEs, whereas, it is 36 min for pre-CMEs. It indicates that the flares associated with the primary CMEs are also of long duration flares (LDE flares).

The statistical values of mean, median and standard deviation of the properties of pre and primary CMEs are given in Table 4. From this table, it is also clear that the primary CMEs are much more energetic than the pre-CMEs. The flares associated with the pre-CMEs are of short duration with integrated flux nearly seven times less than that of primary-CMEs. Some of the events are found to be associated with major SEP proton events. From GOES proton data, we identified **9** major SEP events (proton flux intensity > 10 pfu) for which the SEP data are taken from the website

**Table 4:** Statistical properties of pre and primary CMEs and their associated flares

| Properties | Pre-CMEs | | | | | Primary CMEs | | | | | | |
|---|---|---|---|---|---|---|---|---|---|---|---|---|
| | Speed (km/s) | Width (deg) | FOD (Rs) | Flare duration (min) | Flare Integrated. Flux (x $10^{-2}$ J/m$^2$) | Speed (km/s) | Width (deg) | FOD (Rs) | Int. Height (Rs) | Time Delay (min) | Flare Duration (min) | Flare Integrated Flux (x $10^{-2}$ J/m$^2$) |
| Mean | 523 | 115 | 11 | 36 | 2 | 1361 | 274 | 22 | 15 | 235 | 55 | 21 |
| Standard Deviation | 440 | 116 | 8 | 28 | 3 | 493 | 113 | 5 | 8 | 167 | 44 | 30 |
| Median | 374 | 59 | 11 | 31 | 1 | 1393 | 360 | 23 | 14 | 211 | 41 | 7 |

www.umbra.nascom.nasa.gov/sep. From the analysis of the flare and SEP data, we found that the SEP intensity is related to the integrated X-ray flux of the flares associated with the primary CMEs. Note that, Ding et al.(2013) reported neither the flare size nor CME speeds are the deciding factors of the size of proton peak intensity of the SEP events. Very recently, Joshi et al.(2013) analyzed extensively a single event on 23 January 2012 and found that the interaction of CMEs have important implications in producing large SEP events. It is also found that all the nine events associated with SEP are western side events and six of them are from northern region. Also for six events, the flares associated with primary CMEs have duration >25 min. The western region has good connectivity with the Earth. Gopalswamy et al.(2008) suggested for space weather applications that if a CME originating from the western hemisphere is accompanied by a DH type II burst, there is a high probability that it will produce an SEP event.

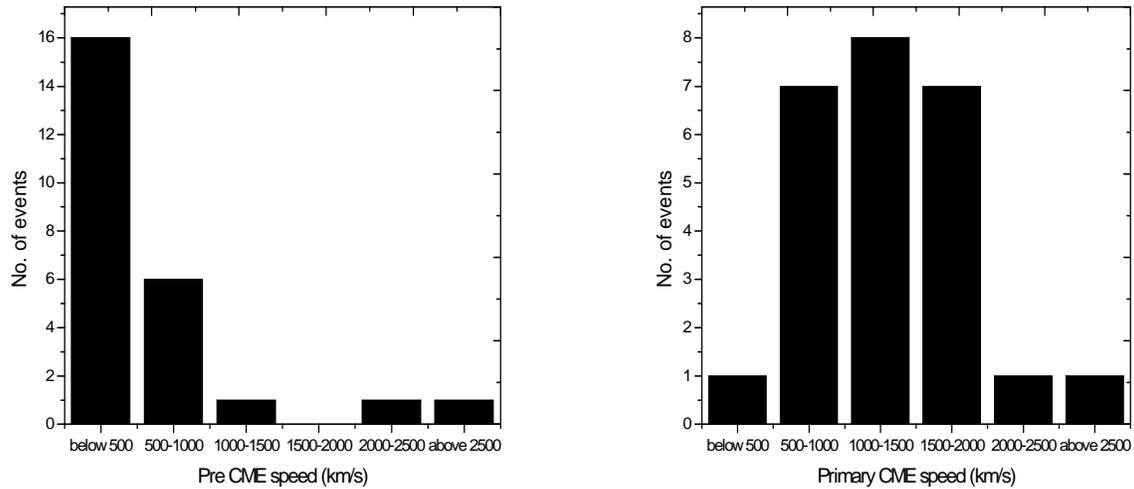

Fig 6 Histogram shows the distribution of speeds of pre and primary CMEs

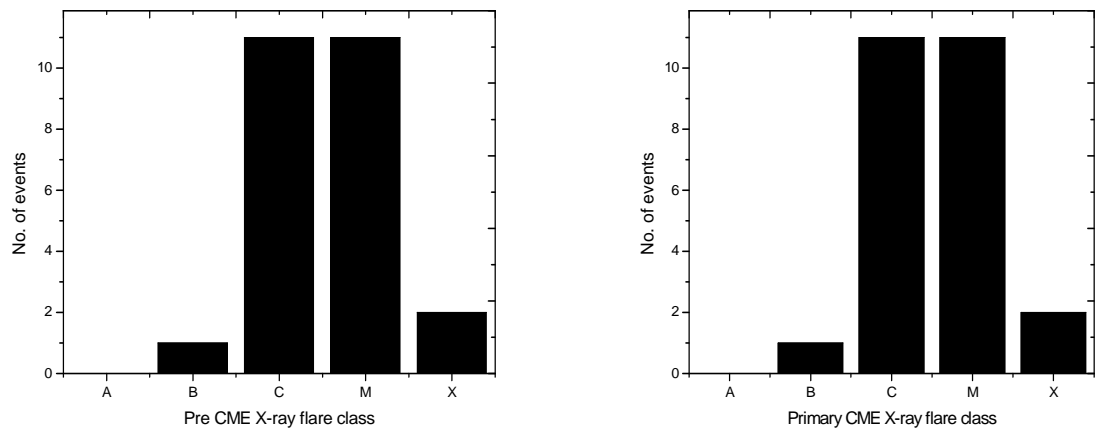

**Fig 7** Histogram shows the distribution of X-ray flare class of pre and primary CMEs.

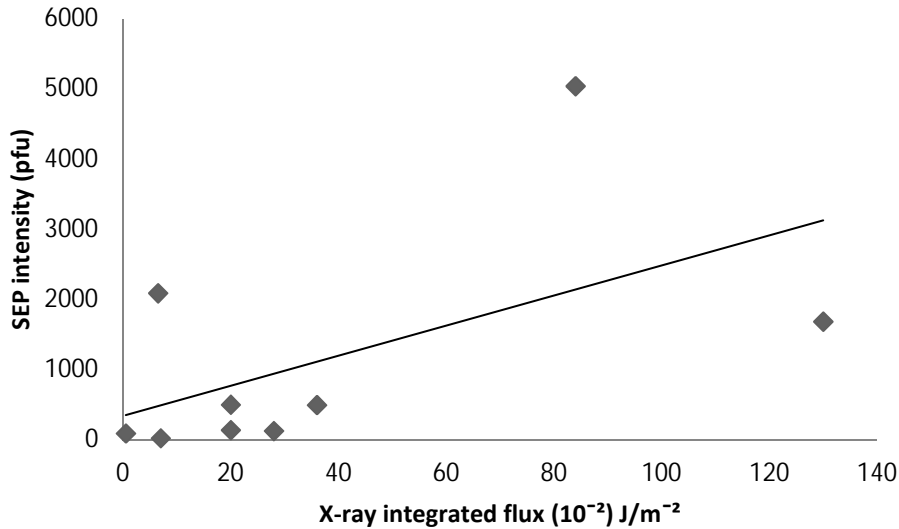

**Fig 8** Distribution of SEP intensity and X ray integrated flux corresponding to the primary CMEs.

## 4 Conclusion

A set of 25 interacting CMEs ejected from the nearby active region in the same quadrant and associated with DH type IIs are identified during the period 1997 – 2010, and their properties are analyzed. These events are collected based on several selection criteria. The pre and primary CMEs and their interactions are identified using LASCO images of SOHO and their height – time diagrams. Most of the interacting events associated with DH type II bursts occurred between the periods 2000-2006 which more number of energetic CMEs were ejected. The associated activities, such as, flares and SEPs are also investigated. Results from this analysis are:

i) Primary CMEs are much faster than the pre-CMEs and their X-ray flares are also stronger (M and X class) compared to the flares (M and C class) of pre-CMEs. The flares associated with primary CMEs are also of longer duration compared to the flares of pre-CMEs.

ii) From the observational data of speed and width of pre and primary CMEs, it is found that the pre-CMEs are found to be less energetic than the primary CMEs.

iii) While the primary CMEs are tracked up to the end of LASCO field of view (30Rs), most of the pre-CMEs can be tracked up to < 26Rs.

iv)The mean values of interaction height and delay between the onsets of pre and primary CMEs are found to be 15 Rs and 235 min.

v) For the 9 events associated with major SEPs, the SEP intensity is found to be related to the integrated X-ray flux of the flares associated with the primary CMEs. These events are from the western region of the Sun.

**Acknowledgement**

We would like to thank Prof. P.K. Manoharan (Radio Astronomy Centre, NCRA-TIFR, Ooty, India) for constructive discussions and the referee for his comments. The authors gratefully acknowledge the data support provided by various online data centers of NOAA and NASA. We would like to thank the Wind/WAVES team for providing the type II catalogs. The CME catalog we have used is provided by the Center for Solar Physics and Space Weather, The Catholic University of America in cooperation with the Naval Research Laboratory and NASA. Also we thank the umbra and NASCOM team for providing Solar Energetic particle data. The major research grant No. 42 – 845 / 2013 (SR) to A.S. from University Grants Commission, Govt. of India is kindly acknowledged.

## References

Ding, L., Jiang, Y., Zhao, L., and Li, G..: Astrophys. J. 763 30 (2013)

Gopalswamy, N., Yashiro, S., Kaiser, M.L., Howard, R.A., Bougeret,J.L.: J. Geophys. Res. 106, 29219 (2001a)

Gopalswamy, N., Yashiro, S., Kaiser, M.L., Howard, R.A., Bougeret,J.L.: Astrophys. J. 548, L91 (2001b)

Gopalswamy,Yashiro S., Michałek,G., Kaiser,M. L, Howard,R. A. Reames,D. V.Leske,R. and Von Rosenvinge,T: Astrophys. J., 572:L103, (2002)

Gopalswamy, N., Aguilar-Rodriguez, E., Yashiro, S., Nunes, S.,Kaiser,M.L., Howard, R.A.: J. Geophys. Res. 110, 12S07 (2005)

Gopalswamy, N., Yashiro, S., Akiyama, S., Makela, P., Xie, H., Kaiser,M.L., Howard, R.A., Bougeret, J.-L.: Ann. Geophys. 26, 3033(2008)

Harrison, R.: Astron. Astrophys., 162, 283(1986)

Harrison, R.: Astron. Astrophys., 304, 585(1995)

Jing, J., et al. : Astrophys. J .620, 1085(2005)

Joshi, N.C. et al. : Adv. Space Res. 52, 1 (2013)

Liu, Y.D., Luhmann, J.G., Christian, M.: The Astrophys. J. Lett., 746, 2, L15, 7(2012).

Lugaz N., Manchester, W.B., Gombosi, T.I.: Astrophys. J.634, 651(2005)

Manoharan, P.K., Gopalswamy, N., Yashiro, S., Lara, A., Howard,R.A., Michalek, G.: J. Geophys. Res. 109, A06109 (2004)

Martínez, O., Juan, C., Raftery, C.L., Bain, H.M., Liu, Y., Krupar, V.,Bale, S., Krucker, S.: Astrophys. J. 748, 66 (2012)

Prasanna Subramanian, S., Shanmugaraju, A. :Astrophys. and Space Sci., 344(2), 305 (2013)

Temmer,M., Vrsnak,B., Rollett, T., et al.: Astrophys. J. 749, 57(2012)

Yashiro, S., Gopalswamy, N., Michalek, G., St. Cyr, O. C., Plunkett,
S. P., Rich, N. B., and Howard, R. A.: J.Geophys. Res., 109, A07105 (2004)

Yashiro, S., Gopalswamy, N.: Universal Heliophysical Processes, eds., 233(2008)

Yan, X.L., Qu, Z.Q., Kong, D.F.: R. Astron. Soc., 414(7), 2803(2011).